# Half-life determination of heavy ions in a storage ring considering feeding and depleting background processes

R. J. Chen[1,2,3,a], G. Leckenby[4,5,b], R. S. Sidhu[1,6,7,c], J. Glorius[1], M. S. Sanjari[1], Yu. A. Litvinov[1,8,9,d], F. C. Akinci[10], M. Bai[1], K. Blaum[3], F. Bosch[1], C. Brandau[1,11], T. Dickel[1,11], I. Dillmann[4], D. Dmytriiev[1], T. Faestermann[12], O. Forstner[1], B. Franczak[1], B. S. Gao[2], H. Geissel[1,11], R. Gernhäuser[12], C. J. Griffin[4], A. Gumberidze[1], E. Haettner[1], R. Heß[1], P.-M. Hillenbrand[1], P. Kienle[12], W. Korten[13], Ch. Kozhuharov[1], N. Kuzminchuk[1], S. Litvinov[1], E. Menz[1], T. Morgenroth[1], C. Nociforo[1], F. Nolden[1], N. Petridis[1], U. Popp[1], S. Purushothaman[1], R. Reifarth[14], C. Scheidenberger[1,11,15], U. Spillmann[1], M. Steck[1], Th. Stöhlker[1,16,17], Y. K. Tanaka[18], M. Trassinelli[19], S. Trotsenko[1], L. Varga[1], M. Wang[2], H. Weick[1], P. J. Woods[6], T. Yamaguchi[20], Y. H. Zhang[2], J. Zhao[1]

[1] GSI Helmholtzzentrum für Schwerionenforschung, Planckstraße 1, 64291 Darmstadt, Germany
[2] CAS Key Laboratory of High Precision Nuclear Spectroscopy, 730000 Lanzhou, China
[3] Max-Planck-Institut für Kernphysik, Saupfercheckweg 1, 69117 Heidelberg, Germany
[4] TRIUMF, 4004 Wesbrook Mall, Vancouver BC V6T 2A3, Canada
[5] Department of Physics and Astronomy, University of British Columbia, Vancouver BC V6T 1Z1, Canada
[6] School of Physics and Astronomy, The University of Edinburgh, Guildford EH9 3FD, UK
[7] School of Mathematics and Physics, University of Surrey, Guildford GU2 7XH, UK
[8] Helmholtz Forschungsakademie Hessen für FAIR (HFHF), GSI Helmholtzzentrum für Schwerionenforschung GmbH, Campus Darmstadt, 64291 Darmstadt, Germany
[9] Institut für Kernphysik, Universität zu Köln, Zülpicher Str. 77, D-50937 Köln, Germany
[10] Department of Physics, Istanbul University, 34134 Istanbul, Turkey
[11] Justus-Liebig Universität Giessen, Leihgesterner Weg 217, 35392 Gießen, Germany
[12] Physik Department E12, Technische Universität München, D-85748 Garching, Germany
[13] IRFU, CEA, Université Paris-Saclay, Gif-sur-Yvette 91191, France
[14] J.W. Goethe Universität, 60438 Frankfurt, Germany
[15] Helmholtz Forschungsakademie Hessen für FAIR (HFHF), GSI Helmholtzzentrum für Schwerionenforschung GmbH, Campus Gießen, 35392 Gießen, Germany
[16] Helmholtz Institute Jena, 07743 Jena, Germany
[17] Institute of Optics and Quantum Electronics, Friedrich Schiller University, 07743 Jena, Germany
[18] High Energy Nuclear Physics Laboratory, RIKEN, Wako, Saitama 351-0198, Japan
[19] Institut des NanoSciences de Paris, CNRS, Sorbonne Université, Paris, France
[20] Saitama University, Saitama 338-8570, Japan

F. Bosch, H. Geissel, P. Kienle and F. Nolden: The author deceased.

[a] e-mail: r.chen@gsi.de (corresponding author)
[b] e-mail: guy.leckenby@gmail.com
[c] e-mail: ragan.sidhu@surrey.ac.uk
[d] e-mail: y.litvinov@gsi.de

**Abstract** Heavy-ion storage rings have relatively large momentum acceptance which allows for multiple ion species to circulate at the same time. This needs to be considered in radioactive decay measurements of highly charged ions, where atomic charge exchange reactions can significantly alter the intensities of parent and daughter ions. In this study, we investigate this effect using the decay curves of ion numbers in the recent $^{205}$Tl$^{81+}$ bound-state beta decay experiment conducted using the Experimental Storage Ring at GSI Darmstadt. To understand the intricate dynamics of ion numbers, we present a set of differential equations that account for various atomic and nuclear reaction processes—bound-state beta decay, atomic electron recombination and capture, and electron ionization. By incorporating appropriate boundary conditions, we develop a set of differential equations that accurately simulate the decay curves of various simultaneously stored ions in the storage ring: $^{205}$Tl$^{81+}$, $^{205}$Pb$^{81+}$, $^{205}$Pb$^{82+}$, $^{200}$Hg$^{79+}$, and $^{200}$Hg$^{80+}$. Through a quantitative comparison between simulations and experi-

Springer

mental data, we provide insights into the detailed reaction mechanisms governing stored heavy ions within the storage ring. Our approach effectively models charge-changing processes, reduces the complexity of the experimental setup, and provides a simpler method for measuring the decay half-lives of highly charged ions in storage rings.

## 1 Introduction

Radioactive decays of highly charged ions are routinely measured in heavy-ion storage rings [1]. Depending on a particular physics case, various experimental techniques can be utilized [2]. For instance, the large acceptance of the storage ring can be used to simultaneously store and measure parent and radioactively decayed daughter ions. This is inevitable in studies of two-body decays, where the difference in orbit lengths of the parent and daughter ions is too small to use particle counters for detection purposes [3,4]. However, the ions continuously interact with the rest gas atoms and—if cooling is required—with electrons of the electron cooler or even with molecules of the internal gas jet target. Therefore, alteration of stored ion intensities via atomic charge exchange processes has to be considered. Also, nuclear decay may have several distinct branches which must be taken into account.

In this work, we present a set of differential equations developed to model and hence describe the decay curves of various ions stored simultaneously within a storage ring [5], based on a recent bound-state beta decay ($\beta_b$) experiment of fully ionized $^{205}$Tl$^{81+}$ ions conducted in the Experimental Storage Ring (ESR) [6,7] at GSI Darmstadt. Using these differential equations, the charge-changing interactions of stable contaminants can be used as calibrants to reduce systematics. Hence, we present a framework to manage the effects of multiple loss mechanisms in storage ring experiments, with a particular focus on decay studies, which can be utilized in future experiments [8–12]. The results of this study are essential to further illustrate the reliability of published results on the $\beta_b$ of $^{205}$Tl$^{81+}$, see [6,7].

## 2 Bound-state beta decay

Before describing the details of the $^{205}$Tl experiment, a brief status of $\beta_b$ measurements is provided. $\beta_b$ is an exotic decay mode in which an electron is directly created in one of the empty atomic orbitals, instead of being emitted into the continuum, and a monoenergetic anti-neutrino is emitted [13,14]. There are two particles in the final state that share the $Q$-value of the decay, making $\beta_b$ decay a two-body process [1,15]. $\beta_b$ decay is the time-reversed analogue of orbital electron capture (EC). The concept of $\beta_b$ decay was initially proposed by Daudel *et al.* in 1947 [13], with the first comprehensive theoretical framework later developed by Bahcall in 1961 [14].

Nearly three decades after these theoretical predictions, $\beta_b$ decay was experimentally observed for the first time in 1992 using the Experimental Storage Ring (ESR) facility [5] at GSI, Darmstadt. This observation involved the $\beta_b$ decay of bare $^{163}_{66}$Dy$^{66+}$ ions (with no electrons) to $^{163}_{67}$Ho$^{66+}$ ions [16]. Using the measured half-life of the $\beta_b$ decay, constraints on the mass of electron neutrino were provided. The measurement campaign was then followed by the $\beta_b$ decay of $^{187}_{75}$Re$^{75+}$ to $^{187}_{76}$Os$^{75+}$ [17]. The $^{187}$Re–$^{187}$Os pair ($\beta^-$ decay; $T_{1/2} = 4.12 \times 10^{10}$ y [18]) is one of the most widely used nuclear clocks, however, its validity as a cosmochronometer was doubted as the half-life of $^{187}$Re depends on its atomic charge state [1]. When fully ionized, $^{187}$Re atoms decay via $\beta_b$ decay with a half-life of $32.0 \pm 2.0$ y [17], which is 9 orders of magnitude lower than the half-life in its neutral state, and hence questioning its application as a cosmochronometer. The next experiment was the first ever direct measurement of the decay rate ratio, $\lambda_{\beta_b}/\lambda_{\beta^-}$, of $\beta_b$ ($^{207}_{81}$Tl$^{81+}$ to $^{207}_{82}$Pb$^{81+}$) to the $\beta^-$ decay ($^{207}_{81}$Tl$^{81+}$ to $^{207}_{82}$Pb$^{82+}$) in $^{207}$Tl$^{81+}$ [19]. The obtained ratio of two- to three-body decay branches are analogous to EC/$\beta^+$ ratios obtained for proton-rich systems [20]. Only two $\lambda_{\beta_b}/\lambda_{\beta^-}$ values have been measured at the ESR so far. In addition to $^{207}$Tl$^{81+}$ [19], both decay branches were measured also in $^{205}_{80}$Hg$^{80+}$ [21].

The most recent measurement is the $\beta_b$ decay measurement of $^{205}_{81}$Tl$^{81+}$ to $^{205}_{82}$Pb$^{81+}$ [6,7]. This measurement has two important physics implications. The first one is linked with the LOREX project (acronym of LORandite EXperiment) [6,22,23], wherein, the measurement is needed to determine the nuclear matrix element of the solar $pp$ neutrino capture by the ground state of $^{205}$Tl to the 2.3 keV excited state in $^{205}$Pb. The second physics case is associated with the $^{205}$Pb–$^{205}$Tl pair as an $s$-process (slow neutron capture process) cosmochronometer [7,17,24].

In a $\beta_b$ decay experiment [6,7,16,17], ions of interest are first created in a nuclear reaction using a primary beam accelerated in a chain of accelerators. These ions are then maximally purified from the numerous other reaction products and finally stored in a high vacuum environment ($\approx 10^{-11}$ mbar) within a storage ring. While stored in the ring, ions and their decay products circulate simultaneously millions of times per second at an energy of 400 MeV/u (about 0.7 times the speed of light). The $\beta_b$ decay constant in the laboratory frame, $\lambda_{\beta_b}$, can be derived from the ratio, $N_D(t_s)/N_P(t_s)$, as a function of the storage time, $t_s$, as follows [25]

$$\frac{N_D(t_s)}{N_P(t_s)} = \lambda_{\beta_b} t_s \left(1 + \frac{1}{2}(\lambda_P^{cc} - \lambda_D^{cc})t_s + ...\right) + \frac{N_D(0)}{N_P(0)} e^{(\lambda_P^{cc} - \lambda_D^{cc})t_s}, \quad (1)$$

where $N_P(t_s)$ and $N_D(t_s)$ are the ion number of the parent and the daughter ions at storage time, $t_s$, respectively. The ratio, $N_D(0)/N_P(0)$, is the initial contamination. $\lambda_P^{cc}$ and $\lambda_D^{cc}$ are the stored beam loss rates (in the laboratory reference frame) due to the atomic charge changing (*cc*) processes in the electron cooler and the residual gas during the storage of parent and daughter ions, respectively. Equation (1) relies on a simplified model that typically includes the $\beta_b$ decay process of the parent nucleus as well as the ion storage losses of both the parent and daughter nuclei.

Typically in such experiments, the beam cooling [26,27] is applied, such that the velocities of all stored particles are the same. The orbits of the ions are then solely defined by the mass-over-charge ratios, $m/q$, of the ions. The cooled parent ions and their $\beta_b$ decay products circulate simultaneously within the ring, where they interact with various media, including residual gases, electrons from the electron cooler, and eventually gas molecules from the gas jet target [28–31]. These interactions can cause the ions to lose or capture[1] electrons, changing their charge states. In turn, ions with altered charge states may also recapture electrons, potentially reverting to their original charge state. Some transmutations may as well lead to ion species whose $m/q$ values correspond to orbits lying outside of the ring acceptance. Such ions are then lost on the walls of the vacuum pipe of the ring. Understanding the dynamics of all these multiple loss mechanisms in a storage ring is crucial for extracting reliable lifetime parameters that are broadly required in storage ring experiments [32–36]. However, the actual experimental cross-feeding process is much more complex than the assumptions underlying this model. Whether this simple model can accurately describe the complex dynamics of the experimental process is important for determining the decay rate in a $\beta_b$ decay experiment.

Furthermore, some parameters in the experiment are difficult to measure directly, such as the storage decay constant of the daughter nucleus $\lambda_D^{cc}$ [6,7,16,17,25]. In many cases, as well as here, the frequencies of parent and daughter ions are mixed and cannot be separated, and thus, a direct measurement of the daughter ions' decay constant is typically not feasible. Instead, it is usually obtained by scaling the storage decay constant of the parent ions. Whether the parameters derived from this scaling method are reasonable is a matter that requires careful examination. Therefore, it is crucial to include these complex processes in a comprehensive framework and further validate the model with experimental data to ensure its accuracy.

---

[1] In the electron cooler, radiative recombination (RR) with free electrons is the primary loss mechanism. However, in the gas target, ion recombination occurs through radiative electron capture (REC) and non-radiative electron capture (NRC) involving bound electrons from the target. Both processes are discussed later in the text.

## 3 $^{205}$Tl experiment

The experimental measurement of the $\beta_b$ decay of the $^{205}$Tl$^{81+}$ ions was performed at the GSI Helmholtzzentrum für Schwerionenforschung in Darmstadt. Since vapors of thallium (Tl) are poisonous, an enriched $^{206}$Pb source was used to produce the beam. Highly charged $^{206}$Pb$^{67+}$ ions were first accelerated to 11.4 MeV/u using the UNILAC linear accelerator and then to 678.46 MeV/u in the heavy-ion synchrotron SIS-18. The $^{206}$Pb$^{67+}$ ions were then impinged on a 1607 mg/cm$^2$ thick beryllium (Be) target backed with 223 mg/cm$^2$ niobium (Nb) to enhance stripping of bound electrons. The target was placed at the entrance of the FRagment Separator (FRS) [37]. $^{205}$Tl$^{81+}$ ions were produced via projectile fragmentation nuclear reactions. The cocktail of fragments was analyzed by the FRS [37], as shown in Figure 1. Having a transmission acceptance in magnetic rigidity $B\rho$ of about $\pm 2\%$ for centered fragments, $B\rho$-$\Delta E$-$B\rho$ separation technique [37] was used in the experiment to extract the purified secondary beams of $^{205}$Tl$^{81+}$ ions. The fragments produced with the same mass-to-charge ratio, $m/q$, were separated with the aid of an energy degrader in the middle focal plane of the FRS. The separation in the energy degrader is based on the principle of the Bethe equation, where the energy loss $\Delta E$ for a given nuclide passing through matter is $\propto Z^2/v^2$, where $v$ and $Z$ are the velocity and charge number of the fragments, respectively. In this way, the bare (fully ionized) $^{205}$Tl$^{81+}$ ions were separated from the nearby H (hydrogen)-like $^{205}$Pb$^{81+}$ ions (same as the $\beta_b$ daughters of $^{205}$Tl$^{81+}$) using an Al degrader with a thickness of 735 mg/cm$^2$. $^{205}$Tl$^{81+}$ ions, along with other ions created in the FRS, were then injected into the ESR [5] at an energy of 400 MeV/u, and the proportion of contaminant $^{205}$Pb$^{81+}$ ions was reduced to only 0.1% of the corresponding bare parent $^{205}$Tl$^{81+}$ ions. To see a significant amount of $\beta_b$-decayed atoms, the accumulation of $^{205}$Tl$^{81+}$ ions up to $\sim 1$–$2 \times 10^6$ was necessary. For this purpose, up to about a hundred bunches of freshly produced $^{205}$Tl$^{81+}$ ions from the FRS were accumulated in the ESR.

In order to reduce the momentum spread in the longitudinal direction and betatron oscillations in the transverse direction of the ions in the ring, stochastic cooling and electron cooling were employed to cool the stored beam [26,27,38]. The stochastic cooling at the ESR is optimized such that the ions with energies around 400 MeV/u are cooled, which is defined by the delay time between the pick-up and the kicker. Therefore, the ions were injected into the ring at 400 MeV/u to enable stochastic cooling. The current of the electron cooling was set to 20 mA and 200 mA during the "storage stage–*A*" and "stripping stage–*B*" (details in the next paragraph), respectively. An argon (Ar) gas jet target was turned on to remove the electron from H-like ions. A 245 MHz non-destructive Schottky resonator detector [39,40] was used to diagnose the beam and measure frequency-resolved ion number inside the ESR [5]. This resonator operates at approximately the 125$^{th}$ harmonic of the revolution frequency, allowing it to convert the frequency spectrum into a mass-to-charge ratio ($m/q$) spectrum of the circulating ions (see Fig. 2). The choice of a high harmonic is pivotal as it enhances the detector's resolving power, enabling precise differentiation of ion species based on their $m/q$ ratios. The detector's fast response time enables it to record a frequency spectrum within just 32 ms, and the intensity of the spectral peaks is directly proportional to the number of particles in the beam. To measure ions produced by recombination and ionization due to atomic charge changing processes in the gas jet target, a multi-wire proportional chamber (MWPC) [41] and a stack of silicon detectors—CsI-Silicon Particle detector for Heavy ions Orbiting in Storage rings (CsISiPHOS) [4]—were used after the first dipole magnet following the gas jet target, respectively. Their positions are shown in Fig. 1.

**Fig. 1** Schematic diagram of the experimental setup, illustrating the key components and equipment used in the experiment

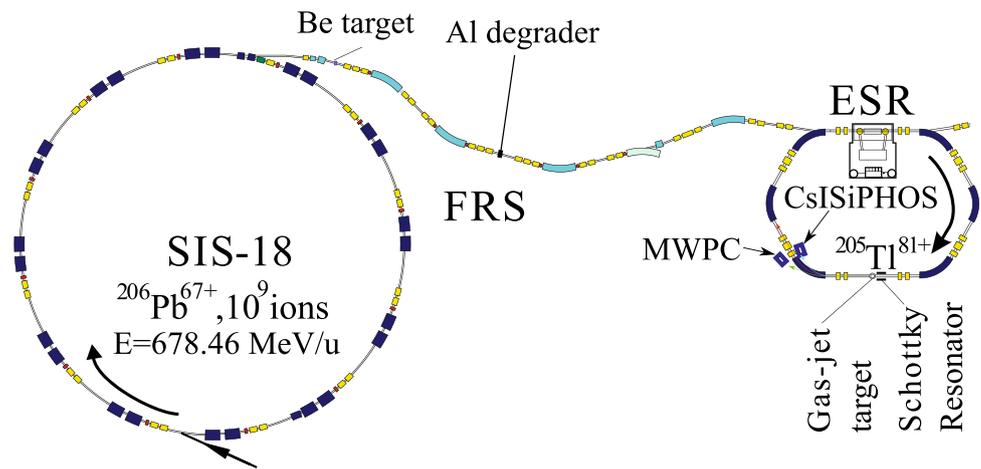

The experimental procedure for measuring the number of $\beta_b$ decay daughter $^{205}$Pb$^{81+}$ ions was as follows (refer to Fig. 2): First, $\approx 1\text{--}2 \times 10^6$ bare $^{205}$Tl$^{81+}$ ions at 400 MeV/u were accumulated and cooled in the ESR for $\approx$ 30–50 minutes (Fig. 2 (a), "accumulation" stage. Second, $^{205}$Tl$^{81+}$ ions were stored for a variable storage time, $t_s$, ranging from 0 to as long as 10 hours (Fig. 2 (a), "storage–A" stage). During these storage times, several hundred $^{205}$Pb$^{81+}$ ions were produced through the $\beta_b$ decay of $^{205}$Tl$^{81+}$. Third, an Ar gas jet target (thickness $\approx 2.2 \times 10^{12}$ atoms/cm$^2$, diameter = 5 mm (FWHM)) was turned on for 600 s to strip off the created $K$-shell electron in $^{205}$Pb$^{81+}$ ions. This step was necessary as $^{205}$Pb$^{81+}$ ions circulated with nearly the same revolution frequency as the $^{205}$Tl$^{81+}$ ions and could not be resolved using the Schottky detector due to the small $Q$-value of the $\beta_b$ decay of only 31.5 keV [42–44] (Fig. 2 (a), "stripping–B" stage). It is to be noted that during the "stripping–B" stage, the observed decrease in $^{205}$Tl$^{81+}$ ions is not due to the stripping process as $^{205}$Tl$^{81+}$ ($Z = 81$) ions are fully ionized and

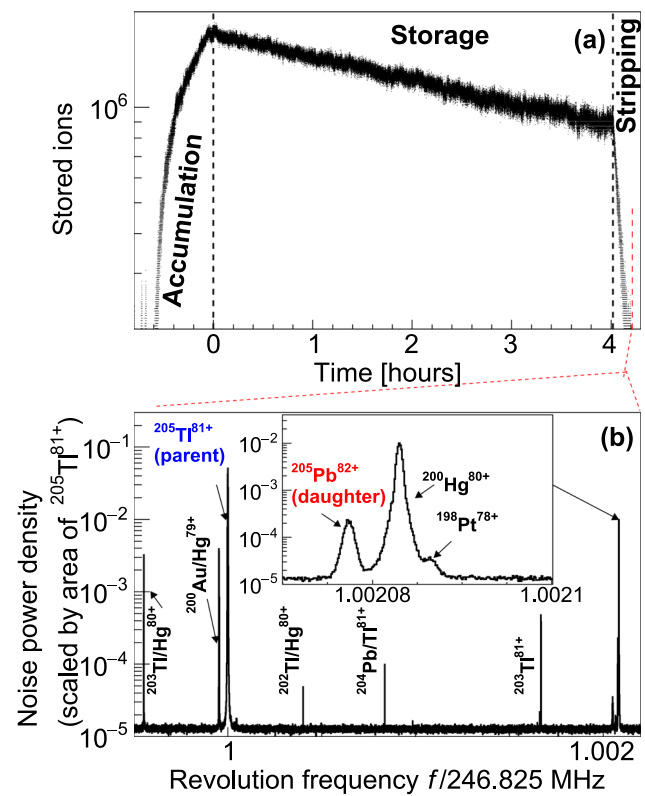

**Fig. 2** (a) The variation of stored ion number with storage time, $t_s$, during a 4 hour storage measurement. (b) The frequency spectrum after a 4 hour storage time and the interaction of the coasting beam with the Ar gas jet target. The inset shows a magnified view of the $^{205}$Pb$^{82+}$ peak relative to that from the nuclear reaction product $^{200}$Hg$^{80+}$ have no electrons to lose. Instead, the reduction results from atomic electron capture from bound electrons in the gas jet target, leading to a charge state transition from $^{205}$Tl$^{81+}$ to $^{205}$Tl$^{80+}$. This process is detailed in Sections 4 and 5. In order to measure the decay constant for the $\beta_b$ decay, both the parent ($^{205}$Tl$^{81+}$) and daughter ($^{205}$Pb$^{82+}$) ions were counted for a time interval of 40 s after 100 s of cooling after the last

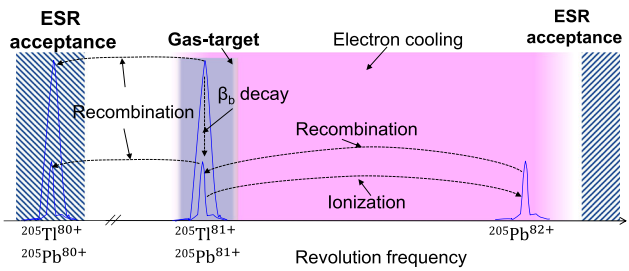

**Fig. 3** Schematic frequency spectrum of products from various atomic charge exchange reactions and nuclear decays of stored ions. The blue diagonal lines at the left and right sides represent the vacuum walls of the ESR. The ions will be lost once they hit the vacuum wall. The violet and pink rectangular areas present the coverage region of the gas jet target and electron cooling. Different decay channels are indicated by arrows

step. More experimental details can be found in references [6,7,25,45].

## 4 Differential equation framework for ion reaction dynamics

In the experiment, $^{205}$Tl$^{81+}$, $^{205}$Pb$^{81+}$, $^{200}$Hg$^{79+}$, and a small number of other ions were created in the FRS and simultaneously stored within the acceptance of the ESR storage ring [5]. Residual $^{205}$Pb$^{81+}$ and $^{200}$Hg$^{79+}$ ions were produced during the nuclear reactions between the primary $^{206}$Pb beam and the Be target, and then transported to the ESR via the FRS [37]. These ions underwent reactions with residual gas in the ring and electrons in the electron cooler, resulting in the production of $^{205}$Tl$^{80+}$, $^{205}$Pb$^{80+}$, $^{205}$Pb$^{82+}$, $^{200}$Hg$^{78+}$, and $^{200}$Hg$^{80+}$ ions either by atomic recombination or ionization processes. Fig. 3 shows the schematic frequency spectrum of $^{205}$Tl$^{81+}$ and $^{205}$Pb$^{81+}$ ions. The schematic diagram for $^{200}$Hg$^{79+}$ ions is similar, except that it does not include the $\beta_b$ decay process.

To describe these processes, we considered several processes affecting the ion populations: $\beta_b$ decay, recombination, and ionization. The $\beta_b$ decay process leads to the conversion of $^{205}$Tl$^{81+}$ ions to $^{205}$Pb$^{81+}$ ions, while recombination can convert $^{205}$Tl$^{81+}$ ions to $^{205}$Tl$^{80+}$ and $^{205}$Pb$^{81+}$ ions to $^{205}$Pb$^{80+}$, and ionization can convert $^{205}$Pb$^{81+}$ ions to $^{205}$Pb$^{82+}$ ions. Conversely, recombination can convert $^{205}$Pb$^{82+}$ ions back to $^{205}$Pb$^{81+}$. The reaction schematics are shown in Fig. 3. The dependence of ion number on storage time, $t_s$, can be described by the following equations

$$\frac{dN_{^{205}Tl^{81+}}(t_s)}{dt_s} = -\lambda_{\beta_b} \times N_{^{205}Tl^{81+}}(t_s) \\ - \lambda_{\text{cap }^{205}Tl^{81+}} \times N_{^{205}Tl^{81+}}(t_s), \quad (2)$$

$$\frac{dN_{^{205}Pb^{81+}}(t_s)}{dt_s} = \lambda_{\beta_b} \times N_{^{205}Tl^{81+}}(t_s) \\ - \lambda_{\text{cap }^{205}Pb^{81+}} \times N_{^{205}Pb^{81+}}(t_s) \\ - \lambda_{\text{str }^{205}Pb^{81+}} \times N_{^{205}Pb^{81+}}(t_s) \\ + \lambda_{\text{cap }^{205}Pb^{82+}} \times N_{^{205}Pb^{82}}(t_s), \quad (3)$$

$$\frac{dN_{^{205}Pb^{82+}}(t_s)}{dt_s} = \lambda_{\text{str }^{205}Pb^{81+}} \times N_{^{205}Pb^{81+}}(t_s) \\ - \lambda_{\text{cap }^{205}Pb^{82+}} \times N_{^{205}Pb^{82+}}(t_s), \quad (4)$$

$$\frac{dN_{^{205}Tl^{80+}}(t_s)}{dt_s} = \lambda_{\text{cap }^{205}Tl^{81+}} \times N_{^{205}Tl^{81+}}(t_s), \quad (5)$$

$$\frac{dN_{^{205}Pb^{80+}}(t_s)}{dt_s} = \lambda_{\text{cap }^{205}Pb^{81+}} \times N_{^{205}Pb^{81+}}(t_s), \quad (6)$$

where $N_{^{205}Tl^{81+}}(t_s)$, $N_{^{205}Pb^{81+}}(t_s)$, $N_{^{205}Pb^{82+}}(t_s)$, $N_{^{205}Tl^{80+}}(t_s)$, and $N_{^{205}Pb^{80+}}(t_s)$ are the ion numbers of $^{205}$Tl$^{81+}$, $^{205}$Pb$^{81+}$, $^{205}$Pb$^{82+}$, $^{205}$Tl$^{80+}$, and $^{205}$Pb$^{80+}$ ions at storage time $t_s$, respectively. $\lambda_{\beta_b}$ is the decay constant of the $\beta_b$ decay of $^{205}$Tl$^{81+}$ ions in the laboratory reference frame. $\lambda_{\text{cap }^{205}Tl^{81+}}$, $\lambda_{\text{cap}^{205}Pb^{81+}}$, and $\lambda_{\text{cap}^{205}Pb^{82+}}$ are the reaction rates of recombination (capture) of $^{205}$Tl$^{81+}$, $^{205}$Pb$^{81+}$, and $^{205}$Pb$^{82+}$ ions, respectively. $\lambda_{\text{str}^{205}Pb^{81+}}$ is the reaction rate of electron ionization (stripping) of $^{205}$Pb$^{81+}$ ions.

The interplay of these processes determines the changes in ion numbers over time. While the equations quantify these changes for each species, the key processes-$\beta_b$ decay, recombination, and ionization-are applicable depending on the electronic charge state of the ions involved.

The boundary conditions for the equations (2)–(6) are the initial ion numbers, $N_{\text{ion}}(0)$, for each ion species ($^{205}$Tl$^{81+}$, $^{205}$Pb$^{81+}$, $^{205}$Pb$^{82+}$, $^{205}$Tl$^{80+}$, and $^{205}$Pb$^{80+}$). The analytical solutions of equations (2)–(6) are provided in Appendix A. The above equations can also be solved numerically using the Python "sympy" package and the code is provided in Appendix B. The numerical results agree well with the analytical results when the same parameters are used.

## 5 Comparison of simulations with experimental data

In this section, we present a detailed comparison between the simulations and experimental data. The comparison focuses on various aspects, including the time delay for daughter ions (see the next paragraph), the estimation of initial parameters during different experimental stages, and the decay constants derived from both storage and stripping stages. Additionally, the section discusses the impact of electron cooling and gas jet interactions on ion loss rates. The parameters used in the simulations are provided in the following subsections.

## 5.1 Time delay for daughter ions

In the experiment, we measured the ion numbers using the Schottky detector and a time delay ($t_d$) was observed for the daughter ions: $^{205}$Pb$^{82+}$, $^{200}$Hg$^{80+}$, and $^{203}$Tl$^{81+}$. This delay occurs because, after their generation, the orbits of the daughter ions differ from those of their corresponding parent ions ($^{205}$Pb$^{81+}$, $^{200}$Hg$^{79+}$, and $^{203}$Tl$^{80+}$) due to the differences in their $m/q$ ratio. While momentum is conserved, the daughter ions no longer maintain the velocity enforced by the electron cooler. Under the influence of electron cooling, the daughter ions gradually adjust their orbits from the initial positions of the parent ions to their final stable positions. Furthermore, the ion-optical dispersion at the target and other locations in the ring is non-zero, which leads to a rigidity mismatch for ions suddenly changing their charge state, leading to increased transverse betatron oscillations. Cooling of transverse motion is known to be less efficient than the longitudinal one [26]. Therefore, the adjustment process takes time, ranging from 0 to 800 seconds, depending on the magnitude of the velocity offset to the nominal velocity. In the above equations (2)–(6), this time delay for the daughter ions is not explicitly included. In order to accurately compare the experimental data with the simulations, it is necessary to shift the simulated ion curves for the daughter ions along the time axis by this delay period. The delay time can be directly obtained from the observed decay curves of the daughter ions. Figure 4 illustrates a typical delay time for the daughter ions $^{200}$Hg$^{80+}$, $^{203}$Tl$^{81+}$, and $^{205}$Pb$^{82+}$.

Due to the limited statistics available for $^{205}$Pb$^{82+}$ ions, it was difficult to precisely determine their delay time. Therefore, we derived the delay times from the better-sampled curves of $^{200}$Hg$^{80+}$ ions and applied these values to $^{205}$Pb$^{82+}$ ions for consistency. Specifically, delay times of 449 seconds during the storage stage and 37 seconds during the stripping stage were introduced for $^{200}$Hg$^{80+}$ ions, and the same delay times were set for $^{205}$Pb$^{82+}$ ions as a baseline. It is important to note that these delay times are specific to the data set shown in Fig. 6. For other measurement sets, slight fluctuations in ion frequencies necessitate recalibration of delay times. Consequently, each data set requires a reassessment of delay times based on the observed curves of $^{200}$Hg$^{80+}$ ions to ensure accurate alignment with experimental observations.

## 5.2 Estimation of initial parameters

During the storage stage, the gas jet target was off and the electron cooling was operated at a low current of 20 mA in order to keep the beam circulating on a specific orbit and minimize ion loss due to radiative recombination (RR) [31], the primary loss mechanism in electron cooling [28–31]. The recombination constant, $\lambda$, of $^{205}$Tl$^{81+}$ ions lost due to capture of electrons in electron cooling,

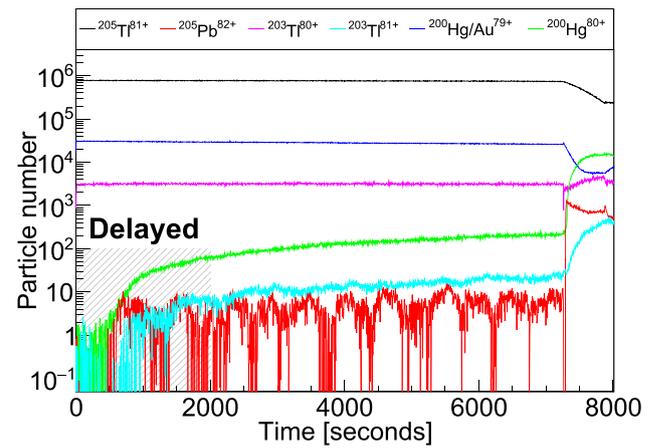

**Fig. 4** Ion number curves as a function of time. The plot presents six distinct curves representing different ion species and their temporal behavior. The black color curve represents $^{205}$Tl/Pb$^{81+}$ ions, red color curve represents $^{205}$Pb$^{82+}$ ions, pink color curve represents $^{203}$Tl/Hg$^{80+}$ ions, light-blue color curve represents $^{203}$Tl$^{81+}$ ions, blue color curve represents $^{200}$Hg/Au$^{79+}$ ions, and green color curve represents $^{200}$Hg$^{80+}$ ions. The ion number was measured by using the Schottky detector

$\lambda^A_{\text{cap}^{205}\text{Tl}^{81+}} = 4.34(6) \times 10^{-5}$ s$^{-1}$, is determined based on the count of $^{205}$Tl$^{81+}$ ions over time measured using a Direct Current Current Transformer (DCCT) [46]. The superscript 'A' indicates that the $\lambda$ is specific to the storage stage.

The recombination constants of $^{205}$Pb$^{81+}$ and $^{205}$Pb$^{82+}$ ions can be obtained by scaling $\lambda^A_{\text{cap}^{205}\text{Tl}^{81+}} = 4.34(6) \times 10^{-5}$ s$^{-1}$ as follows

$$\begin{aligned}
\lambda^A_{\text{cap}^{205}\text{Pb}^{81+}} &= \frac{\text{RR}_{^{205}\text{Pb}^{81+}}}{\text{RR}_{^{205}\text{Tl}^{81+}}} \times \lambda^A_{\text{cap}^{205}\text{Tl}^{81+}} \\
&= \left(\frac{8.3 \times 10^{-5}}{9 \times 10^{-5}}\right) \times 4.34(6) \times 10^{-5} \\
&= 3.99(10) \times 10^{-5} \text{ s}^{-1},
\end{aligned} \quad (7)$$

$$\begin{aligned}
\lambda^A_{\text{cap}^{205}\text{Pb}^{82+}} &= \left(\frac{Z_{^{205}\text{Pb}^{82+}}}{Z_{^{205}\text{Tl}^{81+}}}\right)^2 \times \lambda^A_{\text{cap}^{205}\text{Tl}^{81+}} \\
&= \left(\frac{82}{81}\right)^2 \times 4.34(6) \times 10^{-5} \\
&= 4.45(7) \times 10^{-5} \text{ s}^{-1}.
\end{aligned} \quad (8)$$

Here, RR$_{^{205}\text{Pb}^{81+}}$ and RR$_{^{205}\text{Tl}^{81+}}$ represent the theoretically estimated radiative recombination rates of $^{205}$Pb$^{81+}$ ions and $^{205}$Tl$^{81+}$ ions with electrons from the electron cooler [28], respectively. The ratio, RR$_{^{205}\text{Pb}^{81+}}$/RR$_{^{205}\text{Tl}^{81+}}$, is estimated to be 0.92(2). The superscript 'A' indicates that the recombination constant or stripping constant $\lambda$ is related to the storage stage, as previously defined in the text. Additionally, the recombination constants of $^{200}$Hg$^{79+}$ and $^{200}$Hg$^{80+}$ ions are scaled ($\sigma_{\text{RR}} \propto Z_p^2$ for low collision energy regime [31], where $Z_p$ is the atomic number of the projectile) using

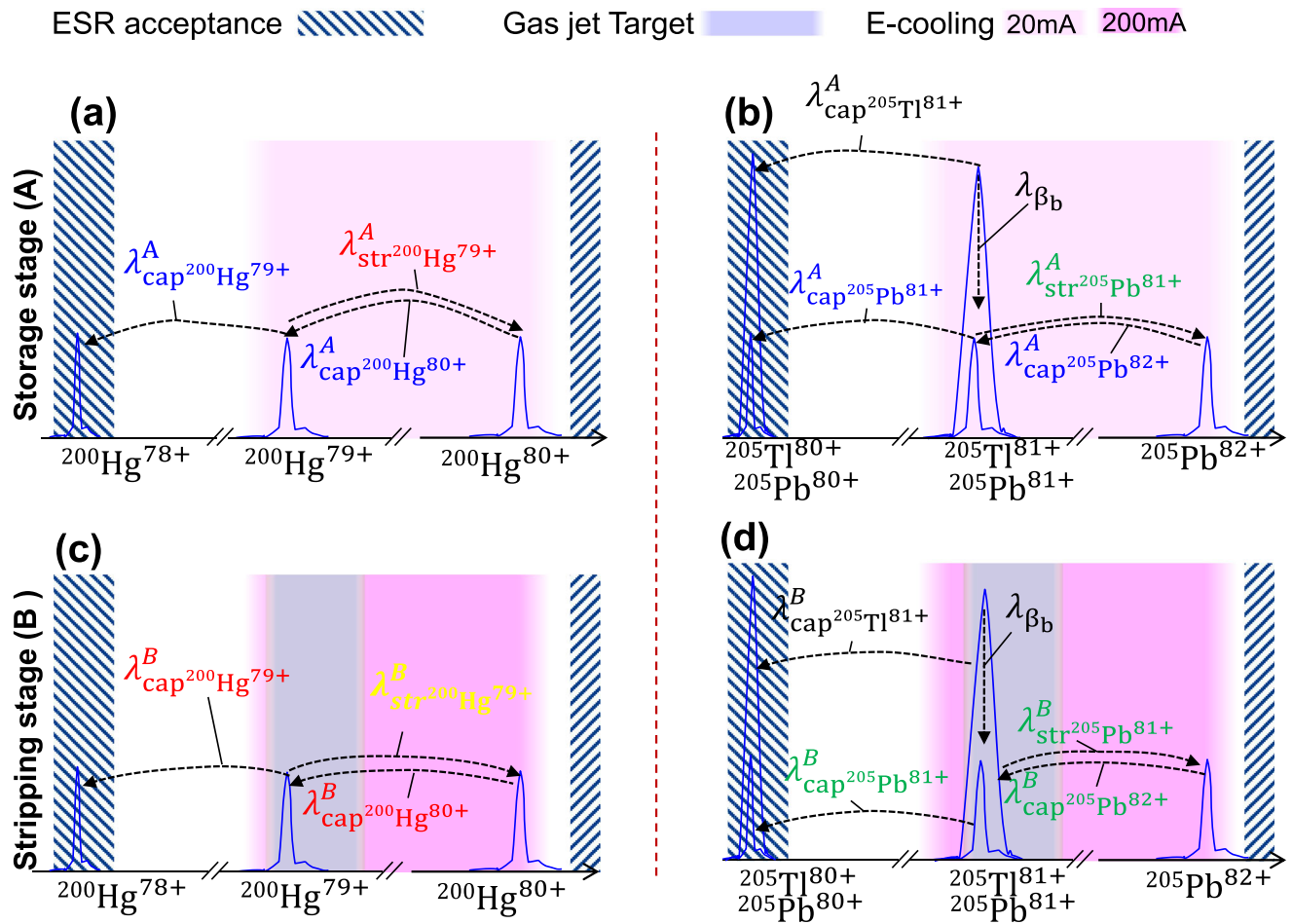

**Fig. 5** Schematic frequency spectrum of products from various atomic charge exchange processes and nuclear decays of stored ions. The left side part **a)** and **c)** is for $^{200}$Hg$^{79+}$ and $^{200}$Hg$^{80+}$ ions, while the right side part **b)** and **d)** is for $^{205}$Tl$^{81+}$ ions, $^{205}$Pb$^{81+}$ ions, and $^{205}$Pb$^{82+}$ ions. The upward part of the illustration represents the storage stage, while the downward part represents the stripping stage. The colors of the rate parameters are explained in the text. The black arrows indicate the processes of ion loss, with the corresponding ion loss rates indicated nearby. The unit of $\lambda$ is s$^{-1}$

$\lambda^A_{\text{cap}^{205}\text{Pb}^{81+}}$ and $\lambda^A_{\text{cap}^{205}\text{Tl}^{81+}}$, respectively, as both ions share the same electronic structure and are given by

$$\lambda^A_{\text{cap}^{200}\text{Hg}^{79+}} = \left(\frac{Z_{^{200}\text{Hg}^{79+}}}{Z_{^{205}\text{Pb}^{81+}}}\right)^2 \lambda^A_{\text{cap}^{205}\text{Pb}^{81+}}$$
$$= \left(\frac{80}{82}\right)^2 \times 3.99(10) \times 10^{-5}$$
$$= 3.80(10) \times 10^{-5} \text{ s}^{-1}, \qquad (9)$$

$$\lambda^A_{\text{cap}^{200}\text{Hg}^{80+}} = \left(\frac{Z_{^{200}\text{Hg}^{80+}}}{Z_{^{205}\text{Tl}^{81+}}}\right)^2 \times \lambda^A_{\text{cap}^{205}\text{Tl}^{81+}}$$
$$= \left(\frac{80}{81}\right)^2 \times 4.34(6) \times 10^{-5}$$
$$= 4.23(6) \times 10^{-5} \text{ s}^{-1}. \qquad (10)$$

During the stripping stage, i.e., when the Ar gas jet target was turned on, the electron cooling was operated with a current of 200 mA. The primary ion loss mechanism is attributed to the interaction of ions with the Ar gas jet target. The atomic electron capture constant, denoted as $\lambda^B_{\text{cap}^{205}\text{Tl}^{81+}}$, represents the decay rate for $^{205}$Tl$^{81+}$ ions lost due to radiative electron capture (REC) and non-radiative electron capture (NRC) with bound electrons from the Ar gas jet target. It is determined based on the ion number over time measured using the MWPC detector. The value of $\lambda^B_{\text{cap}^{205}\text{Tl}^{81+}}$ varied for different measurements due to the fluctuating density of the gas jet target. Additionally, as both the $^{205}$Pb$^{81+}$ and $^{200}$Hg$^{79+}$ ions have similar $m/q$ ratios, their orbits are close to each other, and their atomic electron capture constant ($\sigma_{\text{REC,NRC}} \propto Z_p^5$ [31]) and ionization constant ($\sigma_{\text{str}} \propto Z_p^2$ [47,48]) are scaled accordingly as follows

$$\lambda^{A/B}_{\text{cap}^{205}\text{Pb}^{81+}} = \left(\frac{82}{80}\right)^5 \times \lambda^{A/B}_{\text{cap}^{200}\text{Hg}^{79+}}, \qquad (11)$$

$$\lambda^B_{\text{str}^{205}\text{Pb}^{81+}} = \left(\frac{82}{80}\right)^2 \times \lambda^B_{\text{str}^{200}\text{Hg}^{79+}}, \quad (12)$$

$$\lambda^B_{\text{cap}^{205}\text{Pb}^{82+}} = \left(\frac{82}{80}\right)^5 \times \lambda^B_{\text{cap}^{200}\text{Hg}^{80+}}, \quad (13)$$

$$\lambda^B_{\text{cap}^{200}\text{Hg}^{79+}} = (\text{CR} - 1)\lambda^B_{\text{str}^{200}\text{Hg}^{79+}}, \quad (14)$$

$$\lambda^B_{\text{cap}^{205}\text{Pb}^{81+}} = (\text{CR} - 1)\lambda^B_{\text{str}^{205}\text{Pb}^{81+}}, \quad (15)$$

$$\text{CR} = \frac{\sigma_{\text{cap}^{206}\text{Pb}^{81+}} + \sigma_{\text{str}^{206}\text{Pb}^{81+}}}{\sigma_{\text{str}^{206}\text{Pb}^{81+}}} = 1.425(14). \quad (16)$$

CR represents the ratio of the atomic electron capture (recombination) cross section ($\sigma_{\text{cap}}$) to the electron ionization (stripping) cross section ($\sigma_{\text{str}}$) for the interaction between the H-like $^{205}$Pb$^{81+}$ ions and an Ar gas target. During the experiment, this value was measured using a $^{206}$Pb$^{81+}$ beam.

**Table 1** The various recombination constants and stripping constants ($\lambda$) for atomic electron capture and ionization for $^{200}$Hg and $^{205}$Pb/$^{205}$Tl are listed in the Table. The unit of $\lambda$ is s$^{-1}$

| $^{200}$Hg | $^{205}$Pb/$^{205}$Tl |
|---|---|
| $\lambda^A_{\text{cap}^{200}\text{Hg}^{80+}} = 4.23(6) \times 10^{-5}$ | $\lambda^A_{\text{cap}^{205}\text{Pb}^{82+}} = 4.45(6) \times 10^{-5}$ |
| $\lambda^A_{\text{cap}^{200}\text{Hg}^{79+}} = 3.80(10) \times 10^{-5}$ | $\lambda^A_{\text{cap}^{205}\text{Pb}^{81+}} = 3.99(10) \times 10^{-5}$ |
| $\lambda^A_{\text{str}^{200}\text{Hg}^{79+}} = 1.45(15) \times 10^{-6}$ | $\lambda^A_{\text{str}^{205}\text{Pb}^{81+}} = 1.64(16) \times 10^{-6}$ |
| | $\lambda^A_{\text{cap}^{205}\text{Tl}^{81+}} = 4.34(6) \times 10^{-5}$ |
| $\lambda^B_{\text{cap}^{200}\text{Hg}^{80+}} = 2.15(22) \times 10^{-4}$ | $\lambda^B_{\text{cap}^{205}\text{Pb}^{82+}} = 2.43(24) \times 10^{-4}$ |
| $\lambda^B_{\text{cap}^{200}\text{Hg}^{79+}} = 2.53(25) \times 10^{-3}$ | $\lambda^B_{\text{cap}^{205}\text{Pb}^{81+}} = 2.65(27) \times 10^{-3}$ |
| $\lambda^B_{\text{str}^{200}\text{Hg}^{79+}} = 6.01(60) \times 10^{-3}$ | $\lambda^B_{\text{str}^{205}\text{Pb}^{81+}} = 6.80(68) \times 10^{-3}$ |
| | $\lambda^B_{\text{cap}^{205}\text{Tl}^{81+}} = 3.23(1.4) \times 10^{-3}$ |

### 5.3 Results and discussions

In this experiment, high intensity Schottky signals were saturated [25] due to a mismatch of the amplification factor of the Schottky detector and reference level of the NTCAP system [49]. This led to the inability to directly obtain the number of $^{205}$Tl$^{81+}$ ions from the Schottky signal spectrum. Since the impurity content in the beam was less than 0.1%, the ion number of $^{205}$Tl$^{81+}$ was directly inferred from the DCCT. Low intensity $^{200}$Hg$^{80+}$ and $^{205}$Pb$^{82+}$ ions were precisely identified from the Schottky frequency spectrum (see Fig. 2), and their ion number was calculated from the Schottky frequency spectrum.

The statistics for $^{200}$Hg$^{80+}$ were significantly higher than those for $^{205}$Pb$^{82+}$, and their frequencies were close to each other as shown in Fig. 2. Consequently, we conducted a fitting to the curve of $^{200}$Hg$^{80+}$ to obtain the parameters. The parameters $\lambda^A_{\text{cap}^{200}\text{Hg}^{79+}}$ and $\lambda^A_{\text{cap}^{200}\text{Hg}^{80+}}$ (in blue font, see Fig. 5) are computed using equations (9) and (10), while $\lambda^A_{\text{str}^{200}\text{Hg}^{79+}}$, $\lambda^B_{\text{cap}^{200}\text{Hg}^{79+}}$ and $\lambda^B_{\text{cap}^{200}\text{Hg}^{80+}}$ (in red font) are treated as free parameters. The parameter $\lambda^B_{\text{str}^{200}\text{Hg}^{79+}}$ (in yellow font) is determined based on equation (14). Upon determining the parameters for $^{200}$Hg$^{79+}$ and $^{200}$Hg$^{80+}$, we proceeded to ascertain the $\lambda^{A/B}_{\text{cap}^{205}\text{Pb}^{81+}}$, $\lambda^B_{\text{str}^{205}\text{Pb}^{81+}}$ and $\lambda^B_{\text{cap}^{205}\text{Pb}^{82+}}$ values (in green font) for $^{205}$Pb$^{81+}$ and $^{205}$Pb$^{82+}$ ions during the stripping stage using equations (11)–(13). The final set of parameters is provided in Table 1.

The data presented in Fig. 6 (a) and (b) were obtained using the DCCT detector, while the data in panels (c), (d), (e), and (f) were collected using the Schottky detector. These different data sources introduced additional challenges for accurate error estimation. To address this, we applied the following method for determining the errors: the experimental parameters $\lambda^A_{\text{cap}^{205}\text{Tl}^{81+}}$ and $\lambda^B_{\text{cap}^{205}\text{Tl}^{81+}}$ have errors derived directly from the uncertainties obtained during the data fitting process. The errors for $\lambda^A_{\text{cap}^{205}\text{Pb}^{81+}}$, $\lambda^A_{\text{cap}^{205}\text{Pb}^{82+}}$, $\lambda^A_{\text{cap}^{200}\text{Hg}^{79+}}$, and $\lambda^A_{\text{cap}^{200}\text{Hg}^{80+}}$ were calculated using the error propagation methods as detailed in equations (7)–(10). For the remaining parameters, a uniform systematic error of 10% was applied. More details can be found in [25,50].

The comparison between experimental data and calculations using differential equations, with the corresponding parameters given in Table 1 and Fig. 5, is illustrated in Fig. 6. As evident from Fig. 6, the simulations exhibit a good agreement with the experimental data.

During the storage stage, we obtained $\lambda^A_{\text{str}^{205}\text{Pb}^{81+}} = 1.52(15) \times 10^{-6}$ s$^{-1}$ and $\lambda^A_{\text{cap}^{205}\text{Pb}^{82+}} = 4.45(6) \times 10^{-5}$ s$^{-1}$. The significantly smaller $\lambda^A_{\text{str}^{205}\text{Pb}^{81+}}$ compared to $\lambda^A_{\text{cap}^{205}\text{Pb}^{82+}}$ suggests that only a small portion of the $^{205}$Pb$^{81+}$ ions, initially ionized to the 82+ charge state in the residual gas and electron cooler, recaptured electrons and reverted back to $^{205}$Pb$^{81+}$ ions. Consequently, atomic electron capture stands out as the primary loss process for $^{205}$Pb$^{81+}$ ions during the storage stage, with a beam loss factor of $\lambda^A_{\text{cap}^{205}\text{Pb}^{81+}} = 3.99(10) \times 10^{-5}$ s$^{-1}$. This result shows that the last two terms in equation (3) contribute negligibly and can be ignored. This supports using a simplified model to derive equation (1) for accurately extracting the $\beta_b$ decay constants.

Moving on to the stripping stage, due to the close frequencies of $^{205}$Pb$^{82+}$ and $^{200}$Hg$^{80+}$ ions, the two ion species could not be distinguished when the gas jet target was activated. In order to separate and measure $^{205}$Pb$^{82+}$ ions, the electron cooler was turned on, with cooler current $I = 200$ mA, after the gas jet target was turned off. After some time, the $^{205}$Pb$^{82+}$ ions were cooled and separated in frequency and counted using the Schottky detector. We determined the values of $\lambda^B_{\text{str}^{205}\text{Pb}^{81+}} = 6.32(63) \times 10^{-3}$ s$^{-1}$ and $\lambda^B_{\text{cap}^{205}\text{Pb}^{82+}} = 2.26(23) \times 10^{-4}$ s$^{-1}$. The significantly larger value of $\lambda^B_{\text{str}^{205}\text{Pb}^{81+}}$ compared to $\lambda^B_{\text{cap}^{205}\text{Pb}^{82+}}$ indicates that most of the $^{205}$Pb$^{81+}$ ions were ionized via interactions with the gas jet target. The calculated stripping efficiency is ≈ 97% during the stripping stage.

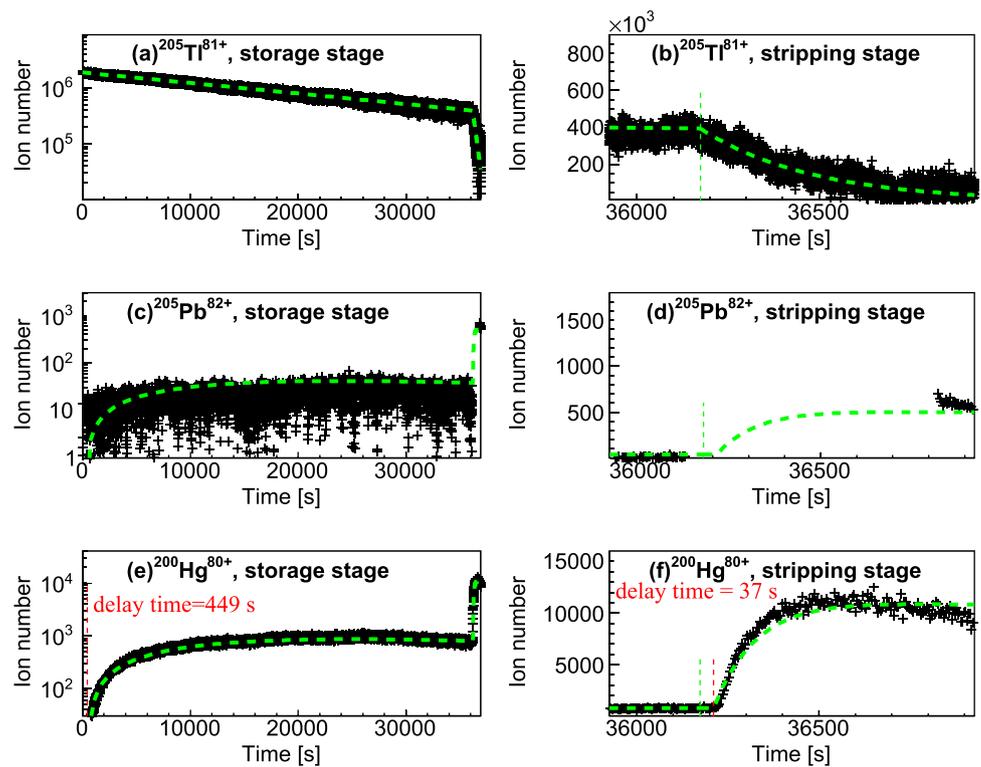

**Fig. 6** This figure presents a comparison between simulations and experimental data for the ions $^{205}$Tl$^{81+}$ (**a**)–(**b**), $^{205}$Pb$^{82+}$ (**c**)–(**d**), and $^{200}$Hg$^{80+}$ (**e**)–(**f**). The black points represent the experimental data collected after a storage time of 10 hours, while the green curves depict the corresponding simulated calculations. The calculations are based on parameters as provided in Table 1. Vertical green dotted lines in (**b**), (**d**), and (**f**) indicate the endpoints of the storage stage, while red dotted lines in (**e**) and (**f**) mark the starting times for the calculation curves. In the calculation curves for $^{205}$Pb$^{82+}$ and $^{200}$Hg$^{80+}$ ions, a delay time of 449 s and 37 s was introduced in the storage and string stages, respectively. Further details are provided in Section 5.1

Furthermore, the value of $\lambda_{\text{cap}^{205}\text{Pb}^{82+}}$ was determined to be $\lambda^{B}_{\text{cap}^{205}\text{Pb}^{82+}} = 2.26(23) \times 10^{-4}$ s$^{-1}$ during the stripping stage, and $\lambda^{A}_{\text{cap}^{205}\text{Pb}^{82+}} = 4.45(6) \times 10^{-5}$ s$^{-1}$ during the storage stage. The ratio of $\lambda^{B}_{\text{cap}^{205}\text{Pb}^{82+}}/\lambda^{A}_{\text{cap}^{205}\text{Pb}^{82+}}$ was ≈ 5.13. Interestingly, this ratio showed the same trend as the ratio of $I$ (electron cooler current) between the two stages, with $I^{B}/I^{A} = (200$ mA / 20 mA$) = 10$. Hence, it is evident that $\lambda_{\text{cap}^{205}\text{Pb}^{82+}}$ is primarily determined by $I$ during the stripping stage. This suggests that the gas jet target did not overlap with the fully ionized ions $^{205}$Pb$^{82+}$ and $^{200}$Hg$^{80+}$, which is an essential conclusion of this work.

Finally, for the determination of the $\beta_b$ decay constant for fully ionized $^{205}$Tl ions, the ratio of the parent ($^{205}$Tl$^{81+}$) and daughter ($^{205}$Pb$^{82+}$) ions was evaluated at the end of the storage stage. Using equation 1, $\beta_b$ decay rate in the center of mass frame was determined to be $\lambda^{\text{c.m.}}_{\beta_b} = \gamma \lambda_{\beta_b} = 2.76(28) \times 10^{-8}$ s$^{-1}$, where $\gamma$ is the Lorentz factor of stored ions. This corresponds to a half-life of $T_{1/2} = 291^{+33}_{-27}$ days. More details on the data analysis can be found in [6,7,25,45,50]. It is interesting to note that the $^{205}$Tl atom, which is stable in a neutral state, becomes radioactive when fully stripped of all electrons. The obtained result has significant implications for the LOREX project and the validity of the $^{205}$Pb–$^{205}$Tl pair as an $s$-process cosmochronometer, which are discussed in detail elsewhere [6,7].

## 6 Conclusions

In this study, we investigated the decay curves of ion numbers of various stored ions in a heavy-ion storage ring. For this purpose, we utilized experimental data from the recent $^{205}$Tl$^{81+}$ bound-state beta decay measurement conducted in the ESR storage ring at GSI, Darmstadt. We developed a set of differential equations to accurately describe the changes in ion numbers, accounting for $\beta_b$ decay, recombination, and electron stripping processes. Additionally, we included the delay in the movement of daughter ions due to electron cooling, leading to an improved match between simulations and experimental data.

Our analysis revealed that the radiative recombination, with electrons of the electron cooler, is the primary factor contributing to the loss of $^{205}$Pb$^{81+}$ ions during storage. Furthermore, we found that nearly 97% of the $^{205}$Pb$^{81+}$ ions were ionized to the 82+ charge state ($^{205}$Pb$^{82+}$) when interacting with the gas jet target during the stripping stage. We also showed that the daughter ions in the 82+ charge state ($^{205}$Pb$^{82+}$) did not overlap with the gas jet targets during this stage, which is an important confirmation of an assumption made during this experiment. This paper provides additional evidence that the parameters used to calculate the $\beta_b$ decay half-life in the experiment are correct.

The results of this work provide a strong framework for modeling the decay rates of the parent and daughter ions. In the past experiments on studies of orbital electron capture [32,33,35,36,51,52], the aperture of the ESR has been dramatically restricted by mechanical scrapers to disable repopulation of the corresponding parent and daughter ions via atomic charge exchange processes. This introduced an additional complication in the experiments since such scraping needs to be technically realized and controlled. The methodology developed here enables us to avoid such scraping and lays a solid foundation for future decay experiments [8–12] at storage ring facilities [53]. This is particularly important for the reliability of the broad-band mass and lifetime measurements planned by the ILIMA (Isomeric Beams, LIfetimes and MAsses) collaboration at FAIR [54].

**Acknowledgements** The results presented here are based on the experiment E121, which was performed using the FRS-ESR facilities in the framework of the SPARC, ILIMA, LOREX, and NucAR collaborations at the GSI Helmholtzzentrum für Schwerionenforschung, Darmstadt (Germany) in the context of FAIR Phase-0 research program. Fruitful discussions and support from G. Amthauer, B. Boev, V. Cvetković, U. Battino, D. Bemmerer, S. Cristallo, R. Grisenti, S. Hagmann, W. F. Henning, A. Karakas, T. Kaur, O. Klepper, W. Kutschera, M. Lestinsky, M. Lugaro, R. Mancino, G. Martínez-Pinedo, B. Meyer, T. Neff, A. Ozawa, V. Pejović, M. Pignatari, D. Schneider, T. Suzuki, B. Szányi, K. Takahashi, S. Yu. Torilov, X. L. Tu, D. Vescovi, P. M. Walker, E. Wiedner, N. Winckler, A. Yagüe Lopéz, X. H. Zhou, and K. Zuber are greatly acknowledged. The authors thank the GSI accelerator team for providing excellent technical support, in particular, to C. Peschke and J. Roßbach from the GSI beam cooling group. We thank the ExtreMe Matter Institute EMMI at GSI, Darmstadt, for support in the framework of an EMMI Rapid Reaction Task Force meeting on the LOREX project. This work was supported by the European Research Council (ERC) under the EU's Horizon 2020 research and innovation programme (Grant Agreement No. 682841 "ASTRUm", Grant Agreement No. 654002 "ENSAR2"); the Natural Sciences and Engineering Research Council of Canada (NSERC) (NSERC Discovery Grant No. SAPIN-2019-00030); the Excellence Cluster ORIGINS from the German Research Foundation DFG (Excellence Strategy EXC-2094-390783311); the State of Hesse within the Research Cluster ELEMENTS (Project ID 500/10.006); and the Science and Technology Facilities Council (STFC) (Grant No. ST/P004008/1). T. Yamaguchi acknowledges support from the Sumitomo Foundation, Mitsubishi Foundation, and JSPS KAKENHI Nos. 26287036, 17H01123, 23KK0055. F.C. Akinci acknowledges support from the Scientific and Technological Research Council of Türkiye 2219 International Postdoctoral Research Fellowship Program for Turkish Citizens (Scholarship No. 1059B191900494), and from Istanbul University Coordination Department of Scientific Research Projects (Project No. FBA-2018-30033). For the purpose of open access, authors have applied a Creative Commons Attribution (CC BY) licence to any Author Accepted Manuscript version arising from this submission.

**Funding** Open Access funding enabled and organized by Projekt DEAL.

**Data Availability Statement** Data will be made available on reasonable request. [Author's comment: The data that support the findings of this study are available from the corresponding author upon reasonable request.]

**Code Availability Statement** Code/software will be made available on reasonable request. [Author's comment: The code used for the simulations and analysis in this study is available from the corresponding author upon reasonable request.]

## A The analytical solutions of equations (2)–(6)

The following equations provide the analytical solutions for the ion number as a function of storage time, based on the decay processes and reactions within the ESR storage ring. The $^{205}$Tl$^{81+}$ ion number as a function of time is given as:

$$N_{^{205}\text{Tl}^{81+}}(t_s) = N_{^{205}\text{Tl}^{81+}}(0) \times \exp\left[-(\lambda_{\text{cap}^{205}\text{Tl}^{81+}} + \lambda_{\beta b, ^{205}\text{Tl}^{81+}}) \times t_s\right]. \quad (17)$$

The $^{205}$Pb$^{81+}$ ion number as a function of time is given as:

$$\begin{aligned}
&N_{^{205}\text{Pb}^{81+}}(t_s) \\
&= E \cdot (a \cdot r_1 \cdot e^{r_1 \cdot t_s} - a \cdot r_2 \cdot e^{r_2 \cdot t_s} + b \cdot e^{r_1 \cdot t_s} - b \cdot e^{r_2 \cdot t_s}) \\
&\quad - \frac{a \cdot D}{r_2 - r_1} \cdot \frac{1}{(-\lambda_{\text{cap}^{205}\text{Tl}^{81+}} - r_1)}(r_1 \cdot e^{r_1 \cdot t_s} \\
&\quad + \lambda_{\text{cap}^{205}\text{Tl}^{81+}} \cdot e^{-\lambda_{\text{cap}^{205}\text{Tl}^{81+}} \cdot t_s}) - \frac{aD}{r_2 - r_1} \\
&\quad \cdot \frac{1}{(-\lambda_{\text{cap}^{205}\text{Tl}^{81+}} - r_2)}(r_2 \cdot e^{r_2 \cdot t_s} + \lambda_{\text{cap}^{205}\text{Tl}^{81+}} \\
&\quad \cdot e^{-\lambda_{\text{cap}^{205}\text{Tl}^{81+}} \cdot t_s}) - \frac{bD}{r_2 - r_1} \cdot \frac{1}{(-\lambda_{\text{cap}^{205}\text{Tl}^{81+}} - r_1)} \\
&\quad \times (e^{r_1 \cdot t_s} - e^{-\lambda_{\text{cap}^{205}\text{Tl}^{81+}} \cdot t_s}) - \frac{bD}{r_2 - r_1} \\
&\quad \cdot \frac{1}{(-\lambda_{\text{cap}^{205}\text{Tl}^{81+}} - r_2)}(e^{r_2 \cdot t_s} - e^{-\lambda_{\text{cap}^{205}\text{Tl}^{81+}} \cdot t_s}). \quad (18)
\end{aligned}$$

The $^{205}\text{Pb}^{82+}$ ion number as a function of time is given as:

$$N_{205\text{Pb}^{82+}}(t_s) = E \cdot e^{r_1 \cdot t_s} - E \cdot e^{r_2 \cdot t_s}$$
$$- \frac{D}{r_2 - r_1} \cdot \frac{1}{(-\lambda_{\text{cap}^{205}\text{Tl}^{81+}} - r_1)}(e^{r_1 \cdot t_s} - e^{-\lambda_{\text{cap}^{205}\text{Tl}^{81+}} \cdot t_s})$$
$$- \frac{D}{r_2 - r_1} \cdot \frac{1}{(-\lambda_{\text{cap}^{205}\text{Tl}^{81+}} - r_2)}(e^{r_2 \cdot t_s} - e^{-\lambda_{\text{cap}^{205}\text{Tl}^{81+}} \cdot t_s}). \quad (19)$$

The $^{205}\text{Tl}^{80+}$ ion number as a function of time is given as:

$$N_{205\text{Tl}^{80+}}(t_s)$$
$$= N_{205\text{Tl}^{80+}}(0) - \frac{\lambda_{\text{cap}^{205}\text{Tl}^{81+}}}{\lambda_{\text{cap}^{205}\text{Tl}^{81+}} + \lambda_{\beta b, 205\text{Tl}^{81+}}} N_{205\text{Tl}^{81+}}(0)$$
$$\times \left\{ \exp\left[-(\lambda_{\text{cap}^{205}\text{Tl}^{81+}} + \lambda_{\beta b, 205\text{Tl}^{81+}}) \cdot t_s\right] - 1 \right\}. \quad (20)$$

The $^{205}\text{Pb}^{80+}$ ion number as a function of time is given as:

$$N_{205\text{Pb}^{80+}}(t_s) = N_{205\text{Pb}^{80+}}(0) + \lambda_{\text{cap}^{205}\text{Pb}^{81+}}$$
$$\left\{ E\left(a \cdot e^{r_1 \cdot t_s} - a \cdot e^{r_2 \cdot t_s} + \frac{b}{r_1} \cdot e^{r_1 \cdot t_s} - \frac{b}{r_2} \cdot e^{r_2 \cdot t_s}\right) \right.$$
$$- \frac{aD}{r_2 - r_1} \cdot \frac{1}{(-\lambda_{\text{cap}^{205}\text{Tl}^{81+}} - r_1)}(e^{r_1 \cdot t_s} - e^{-\lambda_{\text{cap}^{205}\text{Tl}^{81+}} \cdot t_s})$$
$$- \frac{aD}{r_2 - r_1} \cdot \frac{1}{(-\lambda_{\text{cap}^{205}\text{Tl}^{81+}} - r_2)}(e^{r_2 \cdot t_s} - e^{-\lambda_{\text{cap}^{205}\text{Tl}^{81+}} \cdot t_s})$$
$$- \frac{b \cdot D}{r_2 - r_1} \cdot \frac{1}{(-\lambda_{\text{cap}^{205}\text{Tl}^{81+}} - r_1)} \left(\frac{1}{r_1} e^{r_1 \cdot t_s} + \frac{1}{\lambda_{\text{cap}^{205}\text{Tl}^{81+}}} e^{-\lambda_{\text{cap}^{205}\text{Tl}^{81+}} \cdot t_s}\right) - \frac{b \cdot D}{r_2 - r_1} \cdot \frac{1}{(-\lambda_{\text{cap}^{205}\text{Tl}^{81+}} - r_2)}$$
$$\left(\frac{1}{r_2} e^{r_2 \cdot t_s} + \frac{1}{\lambda_{\text{cap}^{205}\text{Tl}^{81+}}} e^{-\lambda_{\text{cap}^{205}\text{Tl}^{81+}} \cdot t_s}\right) \right\} - \lambda_{\text{cap}^{205}\text{Pb}^{81+}}$$
$$\left\{ E\left(\frac{b}{r_1} - \frac{b}{r_2}\right) - \frac{b \cdot D}{r_2 - r_1} \cdot \frac{1}{(-\lambda_{\text{cap}^{205}\text{Tl}^{81+}} - r_1)} \right.$$
$$\left(\frac{1}{r_1} + \frac{1}{\lambda_{\text{cap}^{205}\text{Tl}^{81+}}}\right) - \frac{b \cdot D}{r_2 - r_1} \cdot \frac{1}{(-\lambda_{\text{cap}^{205}\text{Tl}^{81+}} - r_2)}$$
$$\left.\left(\frac{1}{r_2} + \frac{1}{\lambda_{\text{cap}^{205}\text{Tl}^{81+}}}\right)\right\}. \quad (21)$$

The parameters $a$, $b$, $B$, $C$, $D$, $E$ used in the above equations are defined as follows:

$$a = \frac{1}{\lambda_{\text{str}^{205}\text{Pb}^{81+}}}, \quad (22)$$
$$b = \frac{\lambda_{\text{cap}^{205}\text{Pb}^{82+}}}{\lambda_{\text{str}^{205}Pb^{81+}}}, \quad (23)$$
$$B = \lambda_{\text{cap}^{205}\text{Pb}^{81+}} + \lambda_{\text{str}^{205}\text{Pb}^{81+}} + \lambda_{\text{cap}^{205}\text{Pb}^{82+}}, \quad (24)$$
$$C = \lambda_{\text{cap}^{205}\text{Pb}^{81+}} \times \lambda_{\text{cap}^{205}\text{Pb}^{82+}}, \quad (25)$$
$$D = N_{205_{text}Tl^{81+}}(0) \times \lambda_{\beta b} \times \lambda_{\text{str}^{205}\text{Pb}^{81+}}, \quad (26)$$
$$r_1 = \frac{-B + \sqrt{B^2 - 4C}}{2}, \quad (27)$$
$$r_2 = \frac{-B - \sqrt{B^2 - 4C}}{2}, \quad (28)$$
$$E = \frac{N_{205\text{Pb}^{81+}}(0)}{a \cdot (r_1 - r_2)} + \frac{2 \cdot a \cdot D}{a \cdot (r_1 - r_2)^2}. \quad (29)$$

## B Sympy Code for Solving Differential equations

The following code demonstrates how to solve a system of differential equations using the Sympy library in Python. This system represents the decay of ion number within the ESR storage ring.

```
from sympy import symbols, Eq, Function, lambdify
from sympy.solvers.ode.systems import dsolve_system
from ROOT import TCanvas
# Define the functions and symbols
N1, N2, N3, N4, N5 = symbols("N1 N2 N3 N4 N5", cls=Function)
x = symbols("x")

# Define the system of differential equations
eqs = [
    Eq(N1(x).diff(x), -1.93e-8*N1(x) - 4.33e-5*N1(x)),
    Eq(N2(x).diff(x), 1.93e-8*N1(x) - 3.99e-5*N2(x)
       - 1.52e-6*N2(x) + 4.44e-5*N3(x)),
    Eq(N3(x).diff(x), 1.52e-6*N2(x) - 4.44e-5*N3(x)),
    Eq(N4(x).diff(x), 4.33e-5*N1(x)),
    Eq(N5(x).diff(x), 3.99e-5*N2(x))
]

# Solve the system of differential equations
solutions = dsolve_system(eqs, ics={
    N1(0): 2.83e6,
    N2(0): 2.83e6 * 1.73e-3,
    N3(0): 0,
    N4(0): 0,
    N5(0): 0
})

# Prepare for plotting the solutions
funcList = []
canvas = TCanvas("canvas", "canvas", 0, 0, 1500, 250)
canvas.Divide(5, 1)

for i in range(0, 5):
    equation = str(solutions[0][i])
    expression = equation.split(",")[1]
    expression = "(" + expression
    title = "func_N" + str(i + 1)
    func_N = TF1(title, expression, 0, 36000)
    funcList.insert(i, func_N)
    canvas.cd(i + 1)
    funcList[i].Draw("l")

canvas.Draw("")
```

This code *snippet* uses the Sympy library to define and solve a system of ordinary differential equations (ODEs). The equations model the decay and interaction processes of ions in the ESR storage ring. The initial conditions are specified, and the solutions are plotted using the ROOT framework.